# Divergence cleaning for weakly compressible smoothed particle hydrodynamics

G. Fourtakas*[a], R. Vacondio[b] and B. D. Rogers[a]
[a]School of Engineering,
The University of Manchester, Manchester,
M13 9PL, UK
[b]Department of Civil and Environmental Engineering and Architecture,
University of Parma,
Parco Area delle Scienze 181/A, 43124,
Parma, Italy
*email: georgios.fourtakas@manchester.ac.uk

**Abstract**

This paper presents a divergence cleaning formulation for the velocity in the weakly compressible smoothed particle hydrodynamics (SPH) scheme. The proposed hyperbolic/parabolic divergence cleaning, ensures that the velocity divergence, *div*(**u**), is minimised throughout the simulation. The divergence equation is coupled with the momentum conservation equation through a scalar field $\psi$. A parabolic term is added to the time-evolving divergence equation, resulting in a hyperbolic/parabolic form, dissipating acoustic waves with a speed of sound proportional to the local Mach number in order to maximise dissipation of the velocity divergence, preventing unwanted diffusion of the pressure field. The *div*(**u**)-SPH algorithm is implemented in the open-source weakly compressible SPH solver DualSPHysics. The new formulation is validated against a range of challenging 2-D test cases including the Taylor-Green vortices, patch impact test, jet impinging on a surface, and wave impact in a sloshing tank. The results show that the new formulation reduces the divergence in the velocity field by at least one order of magnitude which prevents spurious numerical noise and the formation of unphysical voids. The temporal evolution of the impact pressures shows that the *div*(**u**)-SPH formulation virtually eliminates unwanted acoustic pressure oscillations. Investigation of particle resolution confirms that the new *div*(**u**)-SPH formulation does not reduce the spatial convergence rate.

## 1. Introduction

The smoothed particle hydrodynamics (SPH) numerical scheme was originally invented for astrophysical simulations ([1, 2]), then successfully extended to free-surface flows by Monaghan and Kos [3] introducing the weakly compressible assumption. In recent years SPH has been successfully applied to several different applications characterized by complex interfaces [4-7] and strong impact flows such as Pelton turbines [8, 9], wave energy converters [10, 11] and gearboxes [12]. However, it is well known that the classical weakly compressible SPH (WCSPH) formulation is affected by spurious pressure oscillations which reduce the accuracy of the numerical solution or hinder the numerical stability of the scheme. To overcome this issue different approaches have been proposed in literature, one option is the so-called incompressible SPH (ISPH), in which incompressibility is enforced by adopting a Helmholtz decomposition of the velocity field [13-16] and the pressure is obtained from a pressure Poisson equation. However, ISPH schemes require accurate identification of the free surface to impose an explicit boundary condition, are more susceptible to noise from boundary conditions [17], and they are difficult to parallelize [18], particularly on heterogeneous hardware such as GPUs [19].

An alternative approach to improve the accuracy of the pressure field was based on the idea of adding a diffusion term to the continuity equation [20-24] or adopting a Riemann solver to ensure stability [25-28]. In these schemes, based on the classical WCSPH formulation, the free-surface boundary condition is implicitly satisfied [29] and thus, complex interfaces can be easily described. Regardless of the formulation of the pressure stabilization term, acoustic waves persist in the numerical pressure field in the presence of shocks or discontinuities. This issue frequently arises in applications such as fluid-structure interaction with impact flows, sloshing problems, wave breaking, and fragmentation. To mitigate these acoustic oscillations, Sun *et al.* [30] recently introduced an additional diffusion term in the momentum equation, acting as an artificial bulk viscosity. Similarly, Khayyer *et al.* [31] developed the velocity-divergence error mitigating (VEM) formulation to eliminate the pressure component due to non-zero divergence. Although these formulations were developed independently and from different initial concepts, they are mathematically equivalent as both add a diffusion term in the momentum equation based on the gradient of the velocity divergence.

In this paper, we propose a novel approach to eliminate non-physical acoustic waves in the pressure field, arising from the weakly compressible assumption. This method is inspired by Dedner *et al.* [32], who applied it to magnetohydrodynamic (MHD) equations using a finite volume scheme. A generalized Lagrangian multiplier formulation is proposed by adding an extra equation, to the governing equations of continuity and momentum. This approach has also been successfully applied to the SPH scheme for MHD [33-35]. However, in our work the hyperbolic divergence cleaning is performed on the velocity field rather than the magnetic field, and therefore the time evolution of the additional scalar field $\psi$ is defined based on the velocity divergence. To further enhance the performance of divergence cleaning, a local parabolic diffusion term is added to the extra equation for the scalar field $\psi$ [34]. An early adaptation of hyperbolic cleaning can be found in Tricco and Price [36] as part of their MHD work. For our work, we employ the well-known WCSPH solver DualSPHysics [37], an open-source solver capable of simulating millions of particles using OpenMP and CUDA acceleration. Nevertheless, we will restrict our work on the shared memory parallelisation for brevity.

The paper is organized as follows; In section 2, the governing equations and the discretization of the classical weakly compressible SPH equations is presented, followed by the velocity divergence cleaning formulation in section 3. In section 4 we assess the accuracy and robustness of the new methodology with several test cases in 2-D using periodic unbounded and bounded cases in the presence of a free surface with fragmentation before concluding with section 5.

## 2. The numerical method framework

### 2.1. Governing equations

The governing equations employed in the paper are the compressible Navier-Stokes equations under isothermal conditions. In the Lagrangian formalism the governing equations in a fluid domain $\Omega$ read,

$$\frac{D\rho}{Dt} + \rho \nabla \cdot \mathbf{u} = 0, \tag{2.1}$$

$$\frac{D\mathbf{u}}{Dt} = -\frac{1}{\rho}\nabla p + \nabla \cdot \boldsymbol{\tau} + \mathbf{f} \tag{2.2}$$

where $\mathbf{u}(\mathbf{x}, t)$ and $p(\mathrm{x}, t)$ is the velocity and pressure field respectively for a density $\rho(\mathrm{x}, t)$ at time $t \in \mathbb{R}^+$ with $\boldsymbol{\tau}$ the deviatoric component of the total stress and a forcing field $\mathbf{f}(\mathbf{x}, t)$ with $\mathbf{x} \in \mathbb{R}^d$, for $d = 1, \ldots, 3$. It is noted that the operator $D(\cdot)$ denotes the total derivative.

### 2.2. SPH formalism

In this section, the basic SPH principles are recalled, and it is assumed that the reader is familiar with the numerical scheme. For extensive reviews the reader is directed to [38, 39]. In continuous form, the SPH integral approximation is defined as the convolution of a sufficiently smooth function $f(\mathbf{x})$ with a weighting function $W: \mathbb{R}^d \to \mathbb{R}$ in a domain $\Omega$ which reads,

$$\langle f(\mathbf{x}) \rangle := \int_{\Omega_\mathbf{x}} f(\mathbf{x}')W(\mathbf{x} - \mathbf{x}', h)d\mathbf{x}' \tag{2.3}$$

over $\mathbf{x}' \in \mathbb{R}^d$ defined locally as $\Omega \cap \Omega_\mathbf{x}$ i.e., within a compact support at point $\mathbf{x}$. The characteristic length $h \in \mathbb{R}^+$ defines the length of the compact support $\Omega_\mathbf{x}$ of the weighting function or smoothing kernel. Herein, the brackets $\langle \cdot \rangle$ denote an approximation. Starting from Eq. (2.3) and after some algebra, the gradient takes the following form,

$$\langle \nabla f(\mathbf{x}) \rangle := \int_{\Omega_\mathbf{x}} f(\mathbf{x}')\nabla W(\mathbf{x} - \mathbf{x}', h)d\mathbf{x}' \tag{2.4}$$

The smoothing kernel $W$ at the limit of $h \to 0$ must recover the Dirac delta (or impulse function) function $\delta(\mathbf{x})$, constrained by the identity,

$$\int_{\Omega_\mathbf{x}} W(\mathbf{x}, h)d\boldsymbol{x} = 1 \tag{2.5}$$

A number of compactly supported smoothing kernels have been proposed in the literature such as compactly supported Gaussian kernels and B-splines [40], etc., nevertheless, in this work we employ the Wendland $C_2$ kernel due to its simplicity and low computational requirements but most importantly positive Fourier transform which has shown to prevent pairing instability [41]. The Wendland $C_2$ kernel reads,

$$W(\mathbf{x}, h) = a_d \begin{cases} \left(1 - \frac{|\mathbf{x}|}{h}\right)^4 \left(1 + \frac{|\mathbf{x}|}{h}\right) & x < kh \\ 0 & x \geq kh \end{cases} \tag{2.6}$$

where $k \in \mathbb{N}$ and $a_d = \frac{7}{4\pi h}$ and $a_d = \frac{21}{16\pi h}$, respectively, in a two- and three-dimensional space with $k = 2$.

The discrete approximation of Eq. (2.3) and (2.4) reads,

$$\langle f_i \rangle = \sum_j f_j W_{ij} V_j \tag{2.7}$$

$$\langle \nabla f_i \rangle = \sum_j f_j \nabla W_{ij} V_j \tag{2.8}$$

where $i, j$ is the interpolating and neighbouring nodal points, respectively in the discrete domain $\Omega_{\mathbf{x}}$ or so-called particles within the Lagrangian formalism with $f(\mathbf{x}') = f_j$ and $W(\mathbf{x} - \mathbf{x}', h) = W_{ij}$. The discrete volume associated with a particle is defined as the ratio of constant mass over density $V = \frac{m}{\rho}$. Henceforth, the brackets $\langle \cdot \rangle$ will be dropped for brevity.

### 2.3. Discretisation of the governing equations

Herein, we employ the classical weakly compressible formulation of Monaghan [38] with the addition of a density diffusion term in the continuity equation as presented by Fourtakas *et al.* [24],

$$\frac{D\rho_i}{Dt} = \rho_i \sum_j \mathbf{u}_{ij} \cdot \nabla W_{ij} V_j + \delta h c_i \sum_j \boldsymbol{\xi}_{ij} \cdot \nabla W_{ij} V_j, \quad (2.9)$$

$$\frac{D\mathbf{u}_i}{Dt} = -\sum_j m_j \frac{(P_i + P_j)}{\rho_i \rho_j} \nabla W_{ij} + \langle \nabla \cdot \boldsymbol{\tau} \rangle_i + \mathbf{f_i} \quad (2.10)$$

where $\mathbf{u}_{ij} = \mathbf{u}_i - \mathbf{u}_j$ with the term $\boldsymbol{\xi}_{ij}$ defined as a first order derivative of the hydrostatic density. In the context of the so-called weakly compressible SPH the shear forces are approximated as,

$$\nabla \cdot \boldsymbol{\tau} \cong \nu \Delta \mathbf{u} \quad (2.11)$$

with a kinematic viscosity of $\nu$ and by employing the Morris operator [42] for the Laplacian in Eq. (2.11),

$$\nu \Delta \mathbf{u} = 2\nu \sum_j \frac{V_j}{x_{ij}^2} \mathbf{x}_{ij} \cdot \nabla W_{ij} \mathbf{u}_{ij} \quad (2.12)$$

Further, and to improve the accuracy and stability of the scheme the shifting algorithm of Lind *et al.* [16] is utilized.

### 2.4. Boundary conditions and time integration

The solid wall and open boundary conditions employed follow the work of English *et al.* [43] and Tafuni *et al.* [44], respectively, implemented in the open-source solver DualSPHysics [37]. Both boundary conditions extrapolate physical quantities to the boundary domain $\Gamma$ over the boundary surface $\partial \Gamma$ using a first-order consistent interpolation proposed by Liu and Liu [45]. The prescribed initial conditions for each test case will be discussed in section 4 in a case-to-case basis.

The classical weakly compressible formulation of Monaghan [38] uses explicit time integration which must be at least second order due to the Lagrangian motion of the particles attached to the material derivative which implies,

$$\frac{D\mathbf{x}}{Dt} = \mathbf{u} \quad (2.13)$$

therefore, an explicit symplectic second order time integration scheme is used based on the position Verlet scheme.

The time integration uses a variable time step based on the CFL condition, and a force restriction as follows,

$$\Delta t_f = (min)_i \left( \sqrt{\frac{h}{|F_i|}} \right) \quad (2.14)$$

$$\Delta t_{CFL} = \frac{h}{c_0} \quad (2.15)$$

with

$$\Delta t = C \min(\Delta t_f, \Delta t_{CFL}) \tag{2.16}$$

where $C \in \mathbb{R}^+$ is the Courant number usually set to 0.1 in this work to ensure stability and $c_0 \in \mathbb{R}^+$ the global speed of sound of the system. Eq. (2.15) implies a linear relationship between the time step and $\frac{1}{c_0}$, the importance of this will become apparent in the following section.

### 2.5. Weakly compressible formulation

As the focus of the paper is to obtain a nearly incompressible solution with a Mach number of $Ma = 0.1$ as usually practiced in weakly compressible SPH, the system of Eqs (2.1) & (2.2) is closed using a barotropic equation of state that reads,

$$P = \frac{c_0^2 \rho_0}{\gamma}\left[\left(\frac{\rho}{\rho_0}\right)^\gamma - 1\right] \tag{2.17}$$

where $\rho_0$ is the reference density of the fluid and $\gamma$ is the polytropic index of the fluid usually taken as $\gamma = 7$ for water. Note, that a linear equation of state is recovered by using $\gamma = 1$ which implies,

$$\rho = \rho_0 + \left(\frac{\partial \rho}{\partial P}\right)(p - p_0) + \left(\frac{\partial^2 \rho}{\partial P^2}\right)(p - p_0)^2 + \cdots \tag{2.18}$$

with

$$\frac{1}{c_0^2} = \left(\frac{\partial \rho}{\partial p}\right) \tag{2.19}$$

and $(p - p_0)$ the acoustic (wave) pressure. Neglecting the higher order terms, a pressure wave will travel through the fluid with a celerity,

$$c_0 = \sqrt{\frac{p'}{\rho'}} \tag{2.20}$$

where the prime ($\cdot'$) denotes the acoustic part. For a weakly compressible solution, the pressure variations must be bound by,

$$p < 0.01 c_0^2 \rho_0 \tag{2.21}$$

Consequently, as the $Ma$ number decreases towards an incompressible solution, in Eq. (2.15) the time step decreases linearly with the $Ma$ number deeming the weakly compressible formulation computationally expensive

## 3. Divergence cleaning for weakly compressible flows

In order to retain a reasonable time step withing a compressible formulation and maintain the pseudo incompressibility, it is necessary to smooth (or clean) the velocity divergence field. Herein, the aim is not to recover the incompressible solution but clean the velocity divergence and hence, pressure from acoustic waves which are present in WCSPH on impact flows.

We borrow the idea of divergence cleaning usually encountered in magnetohydrodynamics (MHD) to impose the divergence constraint $\nabla \cdot B = 0$ in the magnetic field. There are alternative ways to impose the divergence constraint as elliptic, parabolic and hyperbolic formulations [32] in the context of MHD. The main disadvantage of elliptic and parabolic divergence cleaning is related to the computational cost involved in the solution of a Poisson equation in the former and the diffusive nature of the latter approach. Dedner *et al.* [32] proposed a hyperbolic constrained approach which was found to alleviate these shortcomings. The hyperbolic constrain approach has been employed in smoothed particle magnetohydrodynamics (SPMHD) [33, 34, 46] to impose the divergence constraint $\nabla \cdot B = 0$. To date, the only attempts weakly compressible hydrodynamics are based on elliptic [31] and parabolic [30] divergence cleaning.

We introduce a scalar field $\psi(x, t)$ in the momentum equation Eq. (2.2) using a gradient term,

$$\frac{D\mathbf{u}}{Dt} = -\frac{1}{\rho}\nabla p + \nu \Delta \mathbf{u} + \mathbf{f} - \nabla \psi \tag{3.1}$$

where $\psi \in \mathbb{R}$ evolves in time using a hyperbolic/parabolic equation,

$$\frac{D\psi}{Dt} = -c_0^2 \nabla \cdot \mathbf{u} - \frac{\psi}{\tau} \tag{3.2}$$

with $\tau$ a decay time scale. Note that, $c_0$ defines the global speed of sound of the system. In SPH formalism, the last term on the RHS of Eq. (3.1) is discretised as,

$$-\nabla \psi = \sum_j \psi_{ij} \nabla W_{ij} V_j, \tag{3.3}$$

and Eq. (3.2) is discretised using,

$$\frac{D\psi_i}{Dt} = c_0^2 \sum_j \mathbf{u}_{ij} \cdot \nabla W_{ij} V_j + \frac{\psi_i}{\tau} \tag{3.4}$$

where the non-conservative gradient operators are used to ensure an accurate interpolation.

The first term of Eq. (3.2) on the RHS gives rise to a purely hyperbolic equation that propagates velocity divergence as acoustic waves away from the source whereas the second term is a parabolic term which decays exponentially with the decay time scale $\tau$. Indeed, by taking the divergence of Eq. (3.1) in the absence of the pressure, viscous and external forcing terms and substituting into Eq. (3.2) with $\varphi = \nabla \cdot \mathbf{u}$ one can obtain a wave equation with a damping term,

$$\frac{\partial^2 \varphi}{\partial t^2} - c_0^2 \Delta \varphi + \frac{1}{\tau}\frac{\partial \varphi}{\partial t} = 0 \tag{3.5}$$

with a propagation velocity of $c_0$ which is the maximum permissible propagation speed based on the CFL criterion. The decay time on the RHS of Eq. (3.5) defines the dissipation rate of $\varphi$ and is usually taken as [33, 34]

$$\frac{1}{\tau} = \frac{\sigma c_0}{\lambda} \tag{3.6}$$

where typically $\sigma \in [0,1]$ is a free parameter and the characteristic length $\lambda = h$ defines how quickly the divergence errors decay, converging as $h \to 0$. The choice of propagation speed $c_0$ requires further investigation. Instead of choosing the numerical speed of sound as is customary in SPMHD, we propose to use in Equation (2.6) a local numerical speed of sound defined as

$$c_i = \kappa \sqrt{\frac{p_i}{\rho_i}} \tag{3.7}$$

and therefore Eq. (3.6) takes the form of,

$$\frac{1}{\tau} = \frac{\sigma c_i}{\lambda} \tag{3.8}$$

provided that the decay strength is dictated locally by $\frac{\partial P_i}{\partial \rho_i}$ instead of the initial numerical speed of sound applied to the global system which in DualSPHysics solver is $c_0 = \kappa\, u_{max}$ based on the expected maximum velocity of the system with $\frac{1}{\kappa} \propto M$. This implies an additional restriction on the time stepping constrain in the form of

$$\Delta t_{div(u)} = \frac{h}{c_i} \tag{3.9}$$

Although at first appearance and additional timestep seems restrictive, our numerical experiments concluded that Eq. (3.9) is restricting the timestep only when strong impacts are present for a few timesteps and is of the same order as the CFL condition. In addition, it guarantees that the divergence errors do not propagate faster than $\frac{1}{\kappa}$. In this work $\sigma = 1$ and κ has the same value as the initial numerical speed of sound defined above unless otherwise stated.

## 4. Results and discussion

The test cases examined in this section are unbounded and bounded cases in the presence of a free surface with strong impact on a solid surface, fragmentation and sloshing of the fluid. The divergence of the velocity is reported for the classical SPH and hyperbolic constrained divergence cleaning SPH denoted as SPH-*div(***u***)* in the figures. Further, results are compared with analytical and known experimental data to estimate the extent to which divergence cleaning affects the numerical results.

### 4.1. Taylor-Green vortices

The Taylor-Green vortices is a periodic test case in a 2-D domain consisting of counter rotating vortices decaying temporally due to viscous dissipation. This test case presents a smooth decaying solution, and it is thus adopted here to assess if the proposed formulation causes additional non-physical dissipation to the numerical scheme in absence of shocks and/or strong pressure waves. The analytical incompressible solution for the velocity and pressure reads,

$$u = -\cos(2\pi x)\sin(2\pi y)e^{\xi t}, \qquad (4.1)$$
$$v = \sin(2\pi x)\cos(2\pi y)e^{\xi t}, \qquad (4.2)$$
$$p = \frac{\rho}{4}(\cos(2\pi x) + \sin(2\pi y))e^{2\xi t}, \qquad (4.3)$$

with $\xi = -\frac{8\pi^2}{Re}$ where the $Re = LU_{max}/\nu$ where $L$ is defined as the horizontal length of the domain and $U_{max}$ is the initial maximum velocity of the system with a kinematic viscosity of $\nu = 0.01$ m²/s. Three different particle resolutions have been used with $L/dx = [50, 100, 200]$ at $Re = 100$. The initial conditions for the velocity can be recovered by setting $t = 0$ in Eqs (4.1)-(4.3).

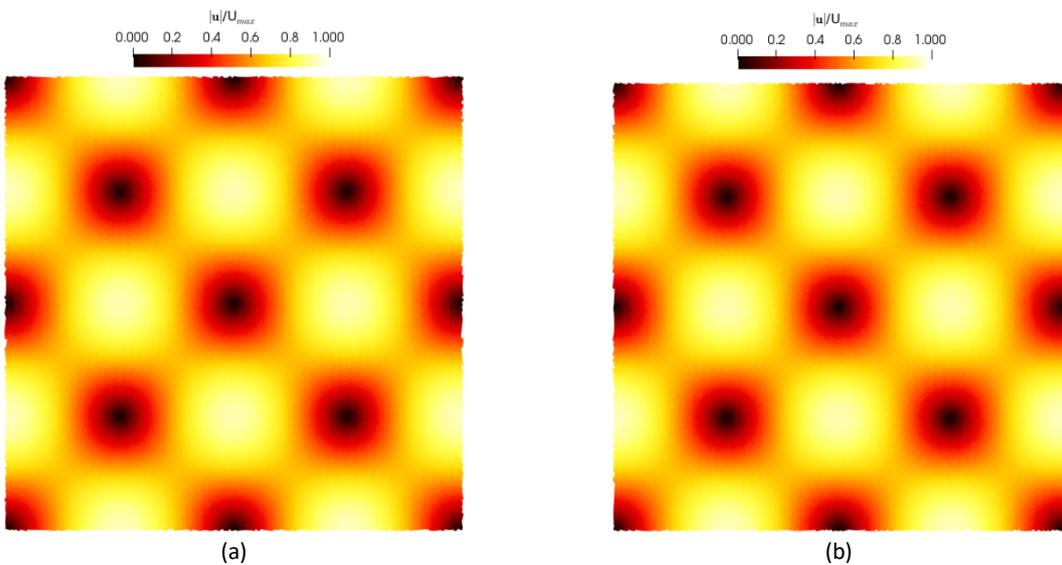

(a) (b)

Figure 1. Taylor-Green vortices velocity snapshot at $t = 2\ s$ (a) with divergence cleaning and (b) without divergence cleaning for $Re = 100$.

Note that, the initial conditions satisfy the divergence free constrain, which makes this case ideal to examine the temporal divergence errors of the velocity for the weakly compressible SPH scheme and the hyperbolic constrained divergence cleaning SPH scheme.

Figure 1 and Figure 2 depict the velocity and pressure field, respectively at $t = 2\ s$ for the resolution of $L/dx = 200$ using the classical SPH formulation of Eqs (2.9) and (2.10) (which we will label as SPH) and the hyperbolic constrained divergence cleaning SPH (which we will refer to as *div(u)-SPH*).

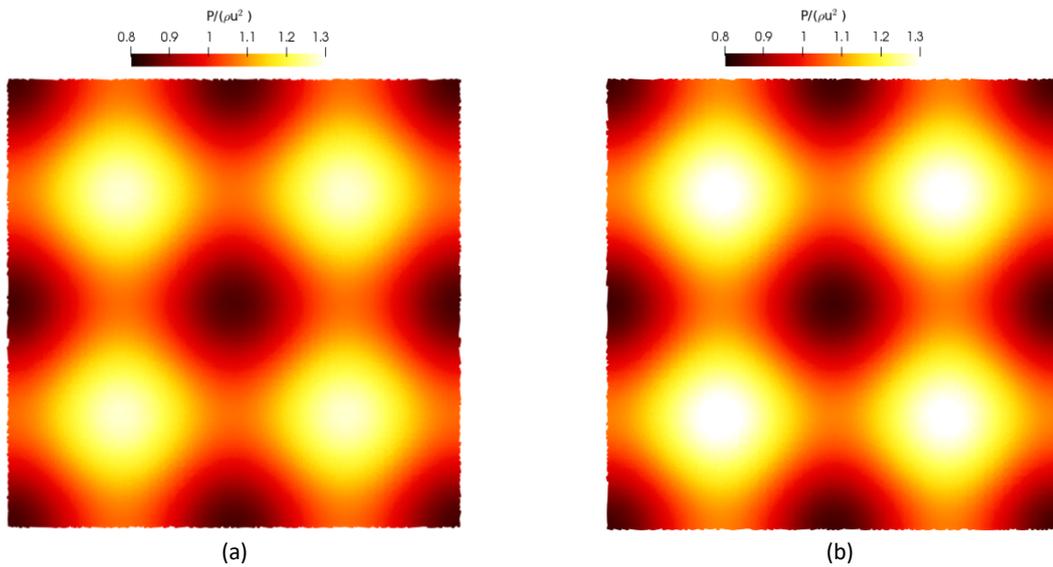

(a)  (b)

Figure 2. Taylor-Green vortices pressure snapshot at $t = 2\ s$ (a) with divergence cleaning and (b) without divergence cleaning for $Re = 100$.

Qualitatively, there are no differences in Figure 1 and Figure 2 in the velocity and pressure fields as the flow field in the Taylor-Green vortices is smooth in the absence of large gradients which will suggest a non-divergence free velocity field. Nevertheless, and despite the incompressible (divergence-free) initial conditions, this is not supported by the normalised *div(u)* flow field depicted in Figure 3.

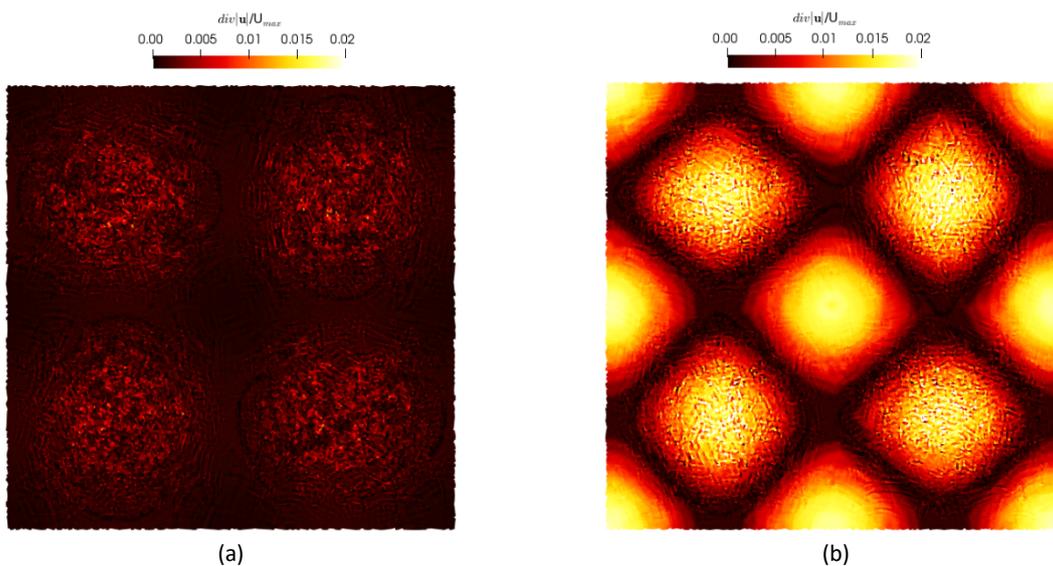

(a)  (b)

Figure 3. Taylor-Green vortices normalised divergence of velocity snapshot at $t = 2\ s$ (a) with divergence cleaning and (b) without divergence cleaning for $Re = 100$.

Figure 3 illustrates the divergence of velocity for the SPH and *div(u)*-SPH scheme at time $t = 2\ s$ for the resolution of $L/dx = 200$. Despite the divergence free initial conditions, the divergence of the classical SPH scheme has drifted towards a compressible solution that indicates larger compressibility than the *div(u)* scheme, indicative of the inability of WCSPH to maintain low levels of velocity divergence. Nevertheless, and despite the qualitative findings of Figure 3, the error characteristics of the scheme remain unchanged as seen in Figure 4 where a convergence analysis has been performed using the $L_2$ velocity error norm.

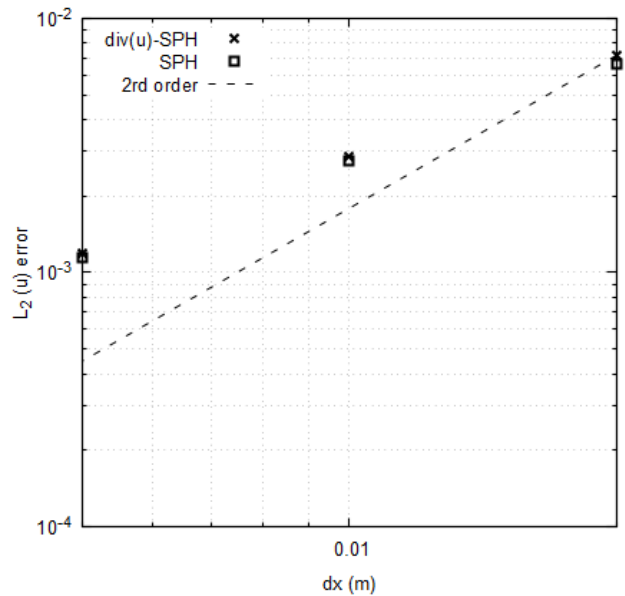

Figure 4. Convergence characteristics for Taylor Green vortices at $t = 4.0$ s and $Re = 100$

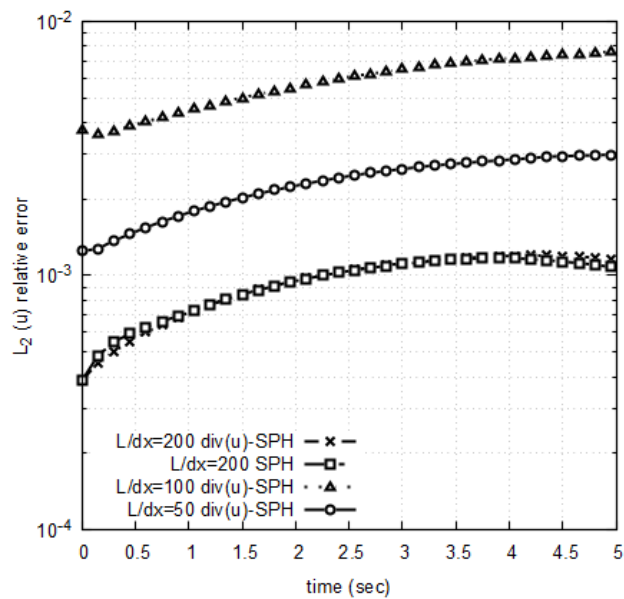

Figure 5. Relative temporal *L₂* error characteristics for *div(u)*- SPH and SPH (without divergence cleaning) for $L/dx = 200$, *div(u)*-SPH for $L/dx = 100$ and $L/dx = 50$.

Both schemes show similar order of convergence of order $O(dx)^{1.4}$ which suggests that the errors due to the compressibility in this case are negligible for a moderate Reynolds number also reinforced by the temporal $L_2$ velocity error norm shown in Figure 5. This is to be expected in a periodic domain with smooth flow field in the absence of a large gradient in the velocity and pressure field.

### 4.2. Patch test

By establishing in section 4.1 that the constrained hyperbolic divergence cleaning procedure proposed does not significantly dissipate the kinematics and dynamics of the flow nor the error characteristics of the SPH scheme, we turn our attention to a free-surface flow that exhibits large deformations and will naturally generate large acoustic pressure waves at impact resulting in spurious acoustic waves. The so-called patch test is a 2-D test case with two identical rectangular patches of liquid that impact at a 0° angle of incidence. Each patch length is $L = 2H$, with a height $H = 1$ m centred around the impact point $\mathbf{x} = [0,0]$, which will serve as the stagnation point. For this case we will use an inviscid fluid with $\rho = 1000$ kg/m³. To maintain stability that arise from numerical errors in SPH, artificial viscosity [3] with $\alpha = 0.01$ is added to the momentum equation using the implementation of DualSPHysics solver [37]. The patch velocity normal to the stagnation point is $u = 1$ m/s. Three different resolutions have been tested with $L/dx = [50,100,200]$ however, we will only discuss the higher resolution as the results do not vary significantly in terms of the velocity divergence errors. Two Mach numbers have been simulated with $Ma = [0.1, 0.05]$.

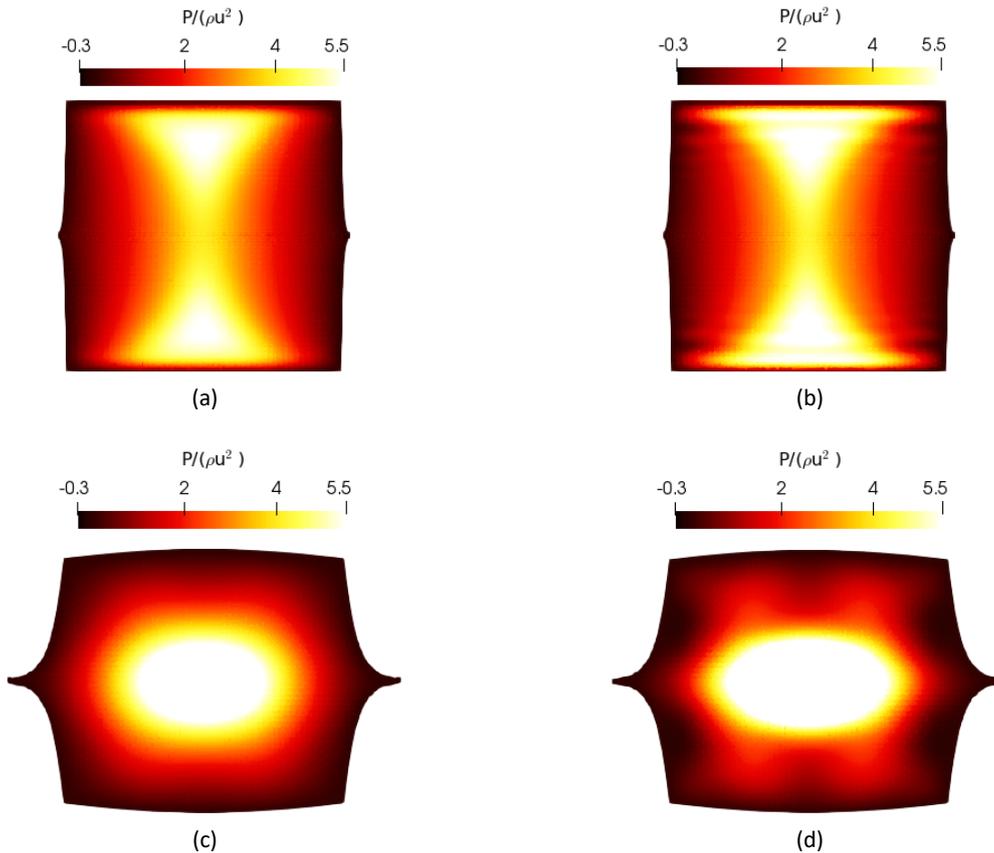

Figure 6. Pressure field at t=0.15 s and t=0.75 s after impact for *div(**u**)-SPH* (a) and (c) and, SPH (b) and (d) for $L/dx = 200$ and $Ma = 0.05$.

Figure 6 shows the pressure field at $t = 0.15$ s and $t = 0.75$ s after impact for the Mach number of $Ma = 0.05$. We choose to show the $Ma = 0.05$ which exhibits less compressibility and thus a smaller amount severe divergence errors to demonstrate the effectiveness of the proposed scheme. The pressure field of the SPH scheme shows spurious acoustic waves propagating from the impact zone to the free surface where they are reflected into the fluid domain.

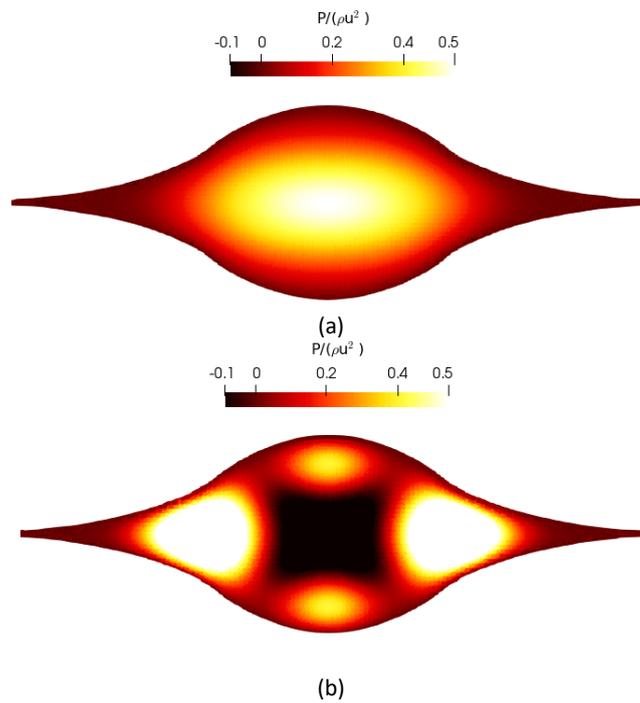

Figure 7. Pressure field snapshots at t=1 s after impact for *div(**u**)-SPH* (a) and SPH (b) for $L/dx = 200$ and $Ma = 0.05$.

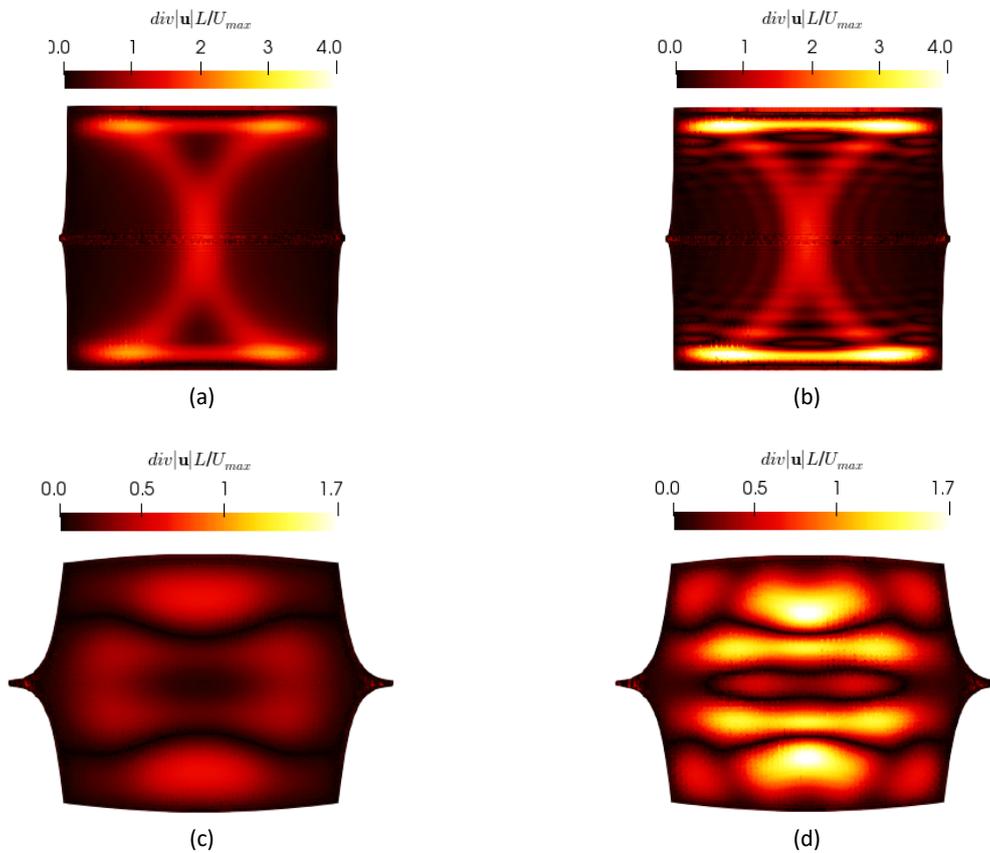

Figure 8. Divergence of velocity snapshots at t=0.15 s and t=0.75 s after impact for *div(**u**)-SPH* (a) and (c) and, SPH (b) and (d) for $L/dx = 200$ and $Ma = 0.05$.

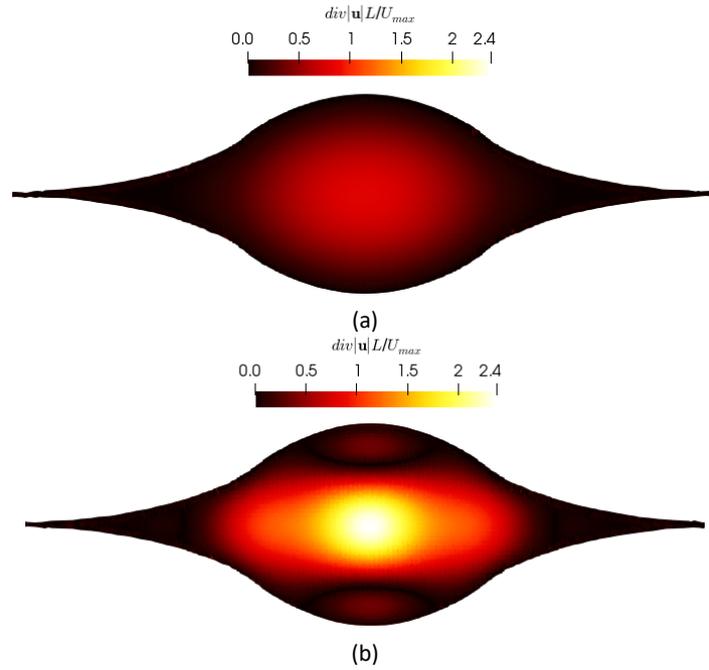

Figure 9. Divergence of velocity snapshots at t=1 s after impact for *div(**u**)-SPH* (a) and SPH (b) for $L/dx = 200$ and $Ma = 0.05$.

This is profoundly evident in Figure 7 where the pressure field oscillates severely due to the acoustic waves at $t = 1$ s. These pressure fluctuations are well known in SPH and are associated with the weak compressibility of the numerical scheme that allows waves to propagate at the numerical speed of sound [47].

It should be noted that similar results were obtained with parabolic divergence cleaning in previous works by Sun *et al.* [30]. Since our aim is to clean the spurious acoustic waves present in the weakly compressible scheme, we turn our attention to the divergence of the velocity in Figure 8 and Figure 9 where we qualitatively compare the velocity divergence for the SPH and *div(**u**)*-SPH schemes.

Figure 8 clearly shows how the acoustic waves propagate through the domain which can be masked in the pressure field as variation may be an of magnitude smaller that the stagnation pressure. Indeed Figure 8 (b), (c) shows that the numerical solution has diverged significantly from the desired $\nabla \cdot \mathbf{u} = 0$ constraint that manifests as density variations through the RHS of Eq. (2.1). Note that, the aim is not to remove the physical pressure field associated with the impact and the weakly compressible approach, but diffuse spurious acoustic wave fluctuations present in the solution due to the impact which is achieved by the parabolic term on the RHS of Eq. (3.2). This is quantified in Figure 10 by plotting the temporal evolution of the pressure at the stagnation point the Mach numbers $Ma = [0.1, 0.05]$. As expected, at the higher Mach number ($Ma = 0.1$) the numerical solution shows that the pressure exhibits larger wavelengths. As the system becomes less compressible in the absence of the divergence constrain, the wavelength decreases, and the wave amplitude increases significantly.

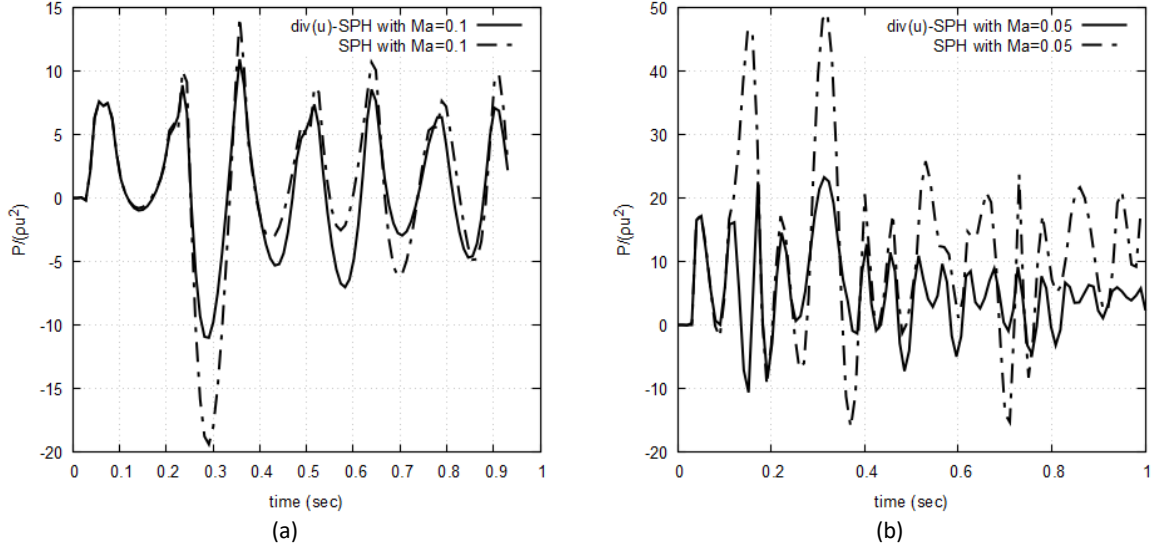

Figure 10. Temporal pressure characteristics at the stagnation point for *div(u)-SPH* and SPH for $L/dx = 200$ with $Ma = 0.1$ and $Ma = 0.05$.

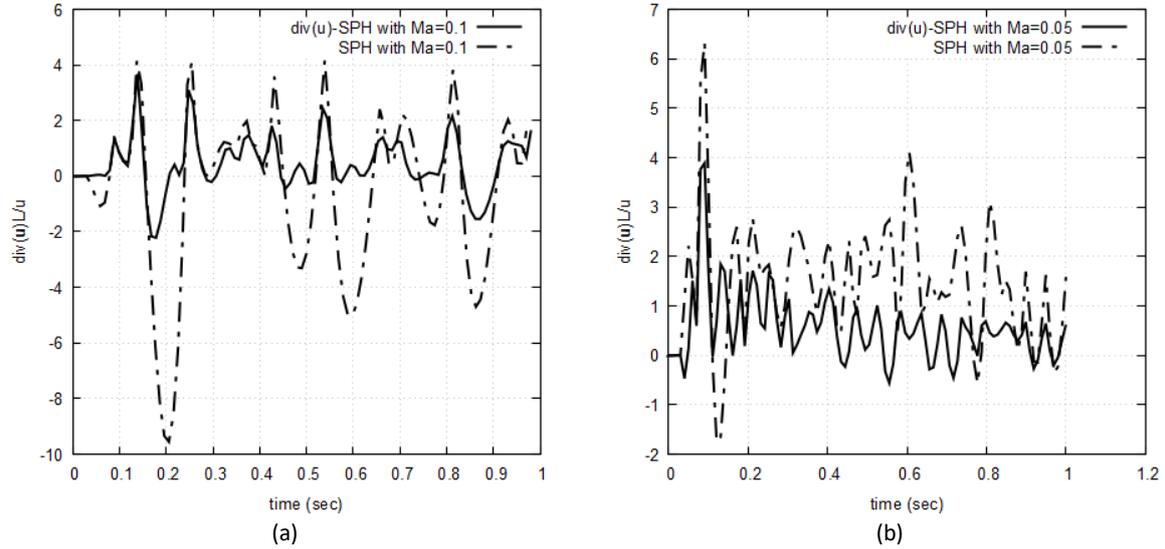

Figure 11. Temporal velocity divergence characteristics for *div(u)-SPH* and SPH for $L/dx = 200$ with $Ma = 0.1$ and $Ma = 0.05$.

Nonetheless, the use of the hyperbolic constrained divergence cleaning reduces the spurious amplitude of the pressure due to the acoustic waves that exaggerate the amplitude. This is also shown in the temporal evolution of the velocity divergence of the system in Figure 11 that shows a reduction near to an order of magnitude in the system.

### 4.3. Jet impinging on a flat surface

An orthogonal 2-D jet impinging on a rigid flat surface is considered. The jet dimensions are defined as $4H = L$ where $H$ is the diameter and $L$ is the length of the jet from the inflow to the flat surface. The outflow conditions of the domain are located at $[-L, 0] \times [L, 0]$ centred around the stagnation point $[0,0]$ with $L = 0.12$ m. The inflow velocity is defined as $U = 20$ m/s with an initial density of $\rho = 1000$ kg/m³. To remain within the weakly compressible regime the speed of sound is chosen to be such that $Ma \approx 0.08$. The kinematic viscosity of the fluid is taken as $\nu = 10^{-6}$ m²/s. The jet diameter is discretised using $H/dx = 61$ particles. Note that Dirichlet boundary conditions are set for Eq. (3.4) with $\psi = 0$ for the wall boundary.

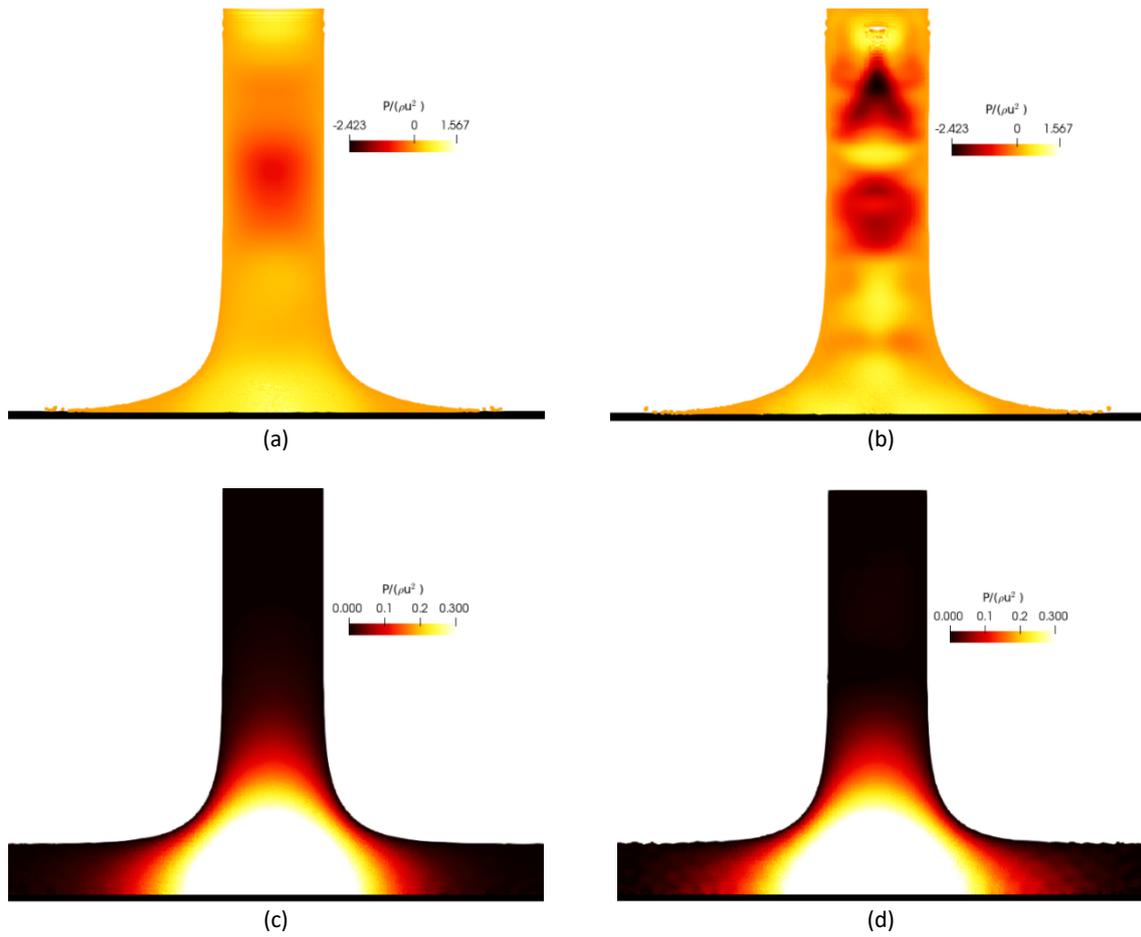

Figure 12. Pressure field at t=0.002 s and t=0.015 s after impact for *div(**u**)-SPH* (a) and (c) and, SPH (b) and (d) for $L/dx = 60$.

Figure 12 shows the pressure field at $t = 0.002$ s and $t = 0.015$ s after the impact of the jet to the surface. As depicted in Figure 12 (a) at $t = 0.002$ s the *div(**u**)*-SPH has reduced the pressure fluctuation significantly at the onset of the impact in contrast to Figure 12 (b) with the classical SPH where pressure waves are propagated and reflected from the inflow boundary condition. It is also notable that the particle distribution near the inflow boundary condition exhibits particle clustering as a strong negative pressure fields bounce through the domain. At $t = 0.015$ s the spurious acoustic waves dissipate for both solutions as the flow reaches steady state. In addition, pressure fluctuations are present away from the stagnation point near the outlet. The spurious acoustic waves are evident in the divergence of the velocity field shown in Figure 13.

In Figure 14, the temporal evolution of the pressure at the stagnation point is plotted. The initial impact pressure of the jet impinging to the surface remains unchanged from the divergence cleaning procedure. However, after the impact the SPH solution shows large fluctuations before the simulation reaches steady state at $t = 0.01$ s. After both solutions have reached steady state, the pressure at the stagnation point exhibits small fluctuations for the SPH simulation which is related to the noise arising from the numerical solution and manifests as spurious pressures near the outflow as shown in Figure 12 and Figure 13 (d). This is confirmed by examining the internal and total energy of the system in Figure 15. At impact there are large fluctuations in the internal energy for the SPH solution which continue to exist in smaller amplitude at steady state and are not pronounced in the *div(**u**)*-SPH internal energy.

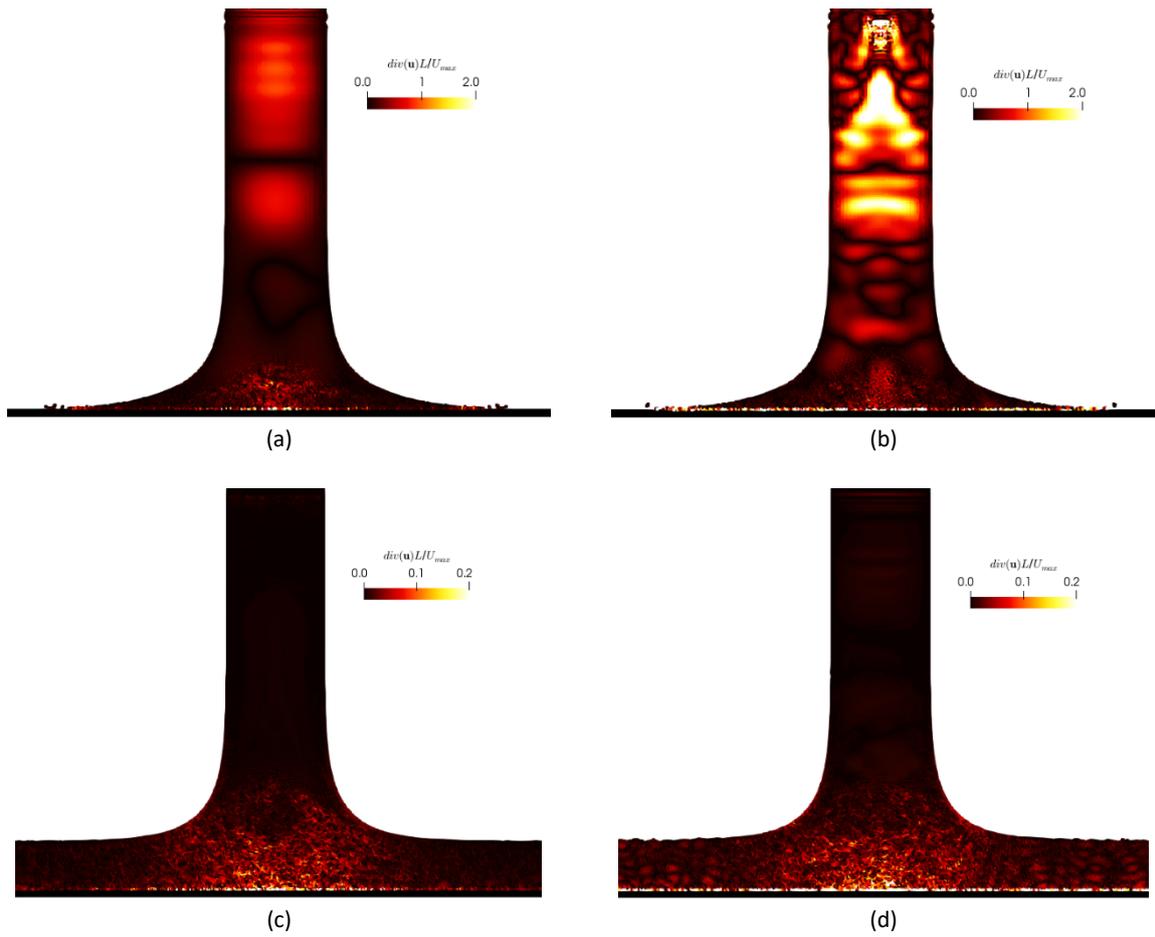

Figure 13. Divergence of velocity snapshots at t=0.002 s and t=0.015 s after impact for *div(**u**)-SPH* (a) and (c) and, SPH (b) and (d) for $D/dx = 60$.

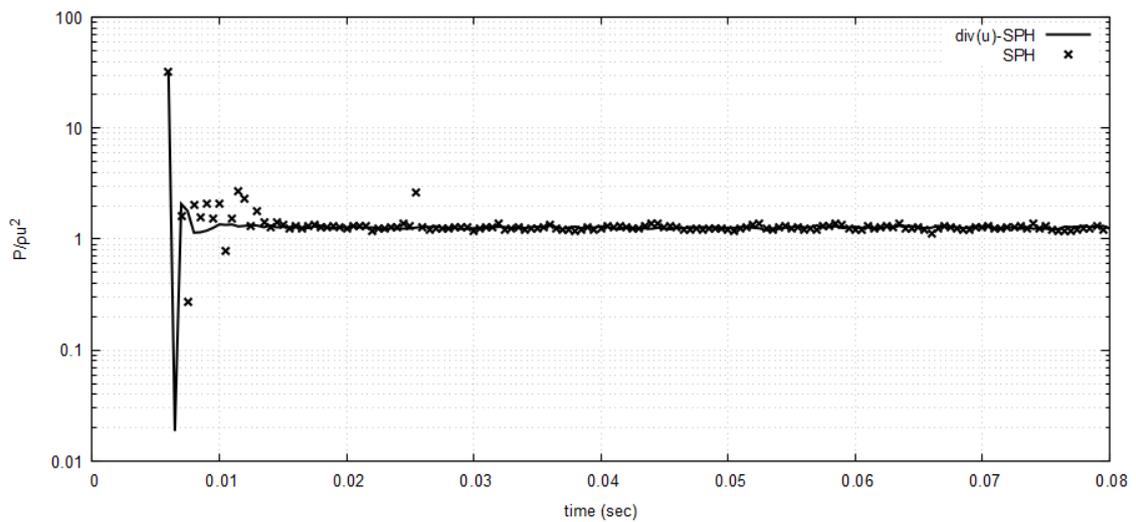

Figure 14. Temporal pressure evolution at the stagnation point for $D/dx = 60$.

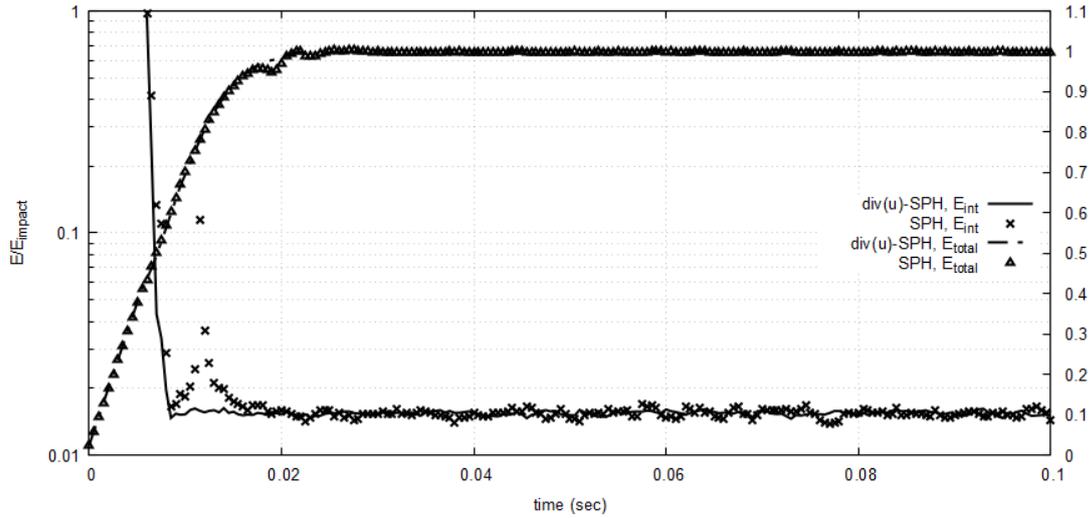

Figure 15. Temporal normalised internal and total energy evolution of the domain for $D/dx = 60$.

### 4.4. Sloshing tank

The final test case to be examined is the sloshing wave impact SPHERIC Benchmark Test Case 10 (https://spheric-sph.org/tests/test-10). A closed rectangular tank with dimensions $[-0.45,0] \times [0.45,0.508]$ has an initial liquid fill level of $H=0.093$ m. A pressure probe is located on the left lateral boundary at the same height as the fill level.

The tank is excited around the rotation axis at the centre of the bottom boundary centred around $[0,0]$ with a prescribed motion. The density of the liquid is $\rho = 1000$ kg/m³. Artificial viscosity with $\alpha = 0.02$ has been used in this test case to guarantee the stability in the momentum equation with a speed of sound $c_0 = 28$ m/s. We use $H/dx = 46$ particles to discretise the liquid fill level.

Figure 16 shows the pressure field at $t = 3.1$ s and $t = 4.9$ s during the first and second impact of the liquid on the left lateral surface of the boundary. At $t = 3.1$ s the improvements on the pressure field using the *div(u)* formulation can be observed. Notably, there are no acoustic pressure waves travelling from the right impact zone to the left. Moreover, the recirculation zone near the right lateral wall can is well formed and prominent. At $t = 4.9$ (second impact) the classical SPH formulation shows large pressure fluctuations propagating from the impact lateral surface. Furthermore, and due to the pressure waves, which oscillate from positive to negative values, a non-physical cavity has formed on the left lateral wall. In contrast, the *div(u)*-SPH does not suffer from spurious noise. The acoustic pressure fluctuations occur in the flow due to the divergence of the velocity as shown in Figure 17 and have very similar structure.

Next, we turn our attention to the temporal evolution of the velocity divergence in Figure 18 where the maximum velocity divergence of the system is shown at first impact at $t = 2.5$ s to avoid contaminating the results by the high frequency oscillations of fragmented flow. It is shown that the maximum divergence of velocity for the *div(u)*-SPH approach is considerably smaller than the SPH.

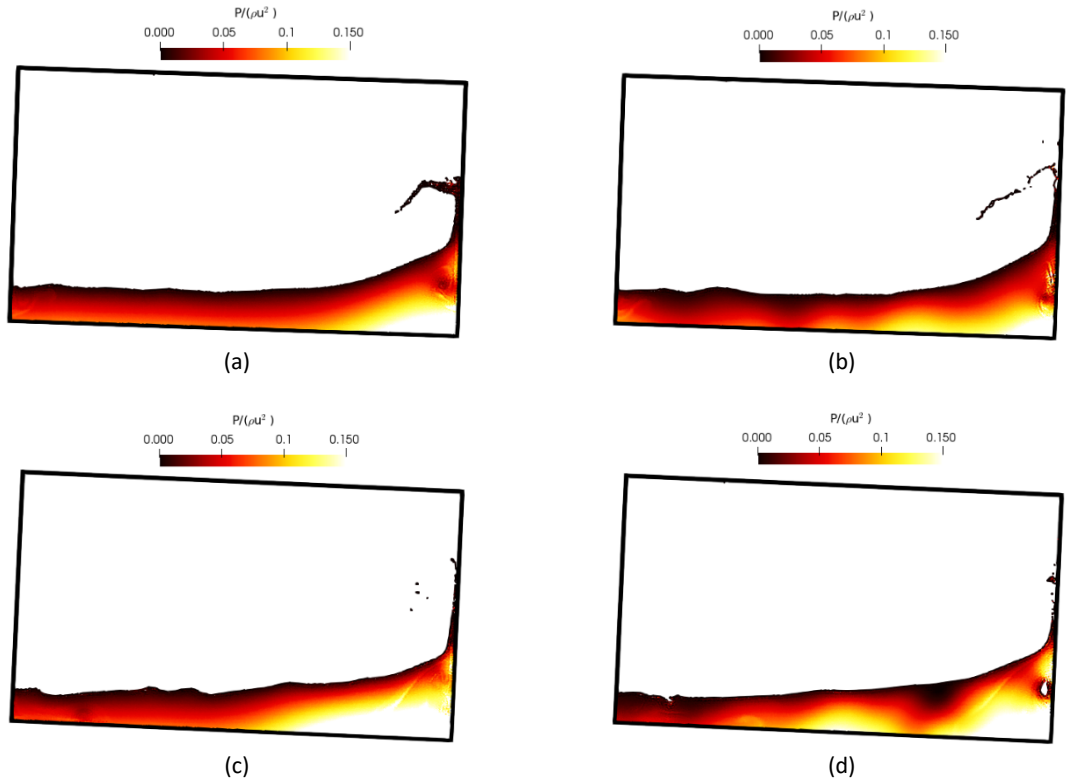

Figure 16. Pressure snapshots at $t = 3.1$ s and $t = 4.9$ s for *div(**u**)-SPH* (a) and (c) and, SPH (b) and (d).

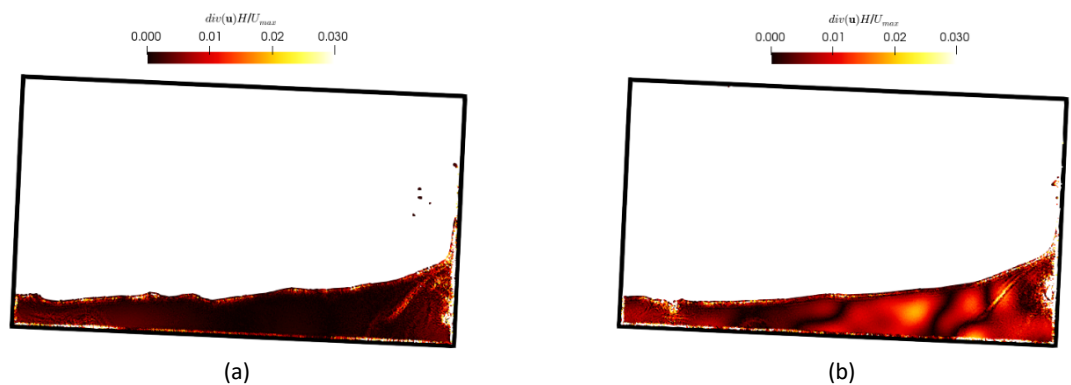

Figure 17. Divergence of velocity snapshots at $t = 4.9$ s for (a) *div(**u**)-SPH* and (b) SPH.

Indeed, Figure 18 shows a reduction of divergence in the velocity near one magnitude. Nevertheless, the authors would like to point out that this is the maximum divergence in the velocity of the fluid domain and contains errors associated with the divergence in the velocity near the boundaries will is naturally not satisfied strictly but is reduced by the *div(**u**)*-SPH approach. By not accounting for particles near the wall boundary where the velocity divergence increases rapidly, the reduction divergence in the velocity is much larger.

In Figure 19 the first two major impacts are plotted. The temporal pressure of *div(**u**)*-SPH exhibits much lower oscillations before and after the impact whereas SPH has a high frequency noise after the impact of the liquid to the lateral left surface. The *div(**u**)*-SPH formulation overestimates the pressure against the experimental data for the pressure probe on the left lateral wall at the first and third impact which could be attributed to the lack of the air phase which cushions the impact and therefore pressures.

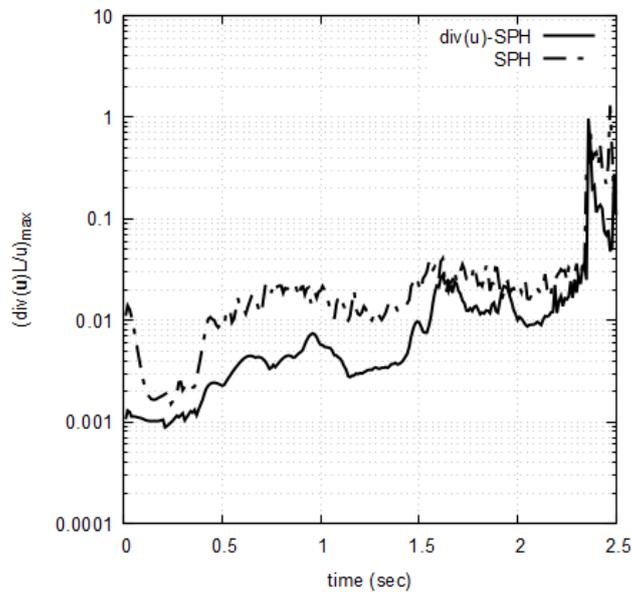

Figure 18. Temporal velocity divergence characteristics for *div(u)-SPH* and SPH for first impact at $t = 2.5$ s.

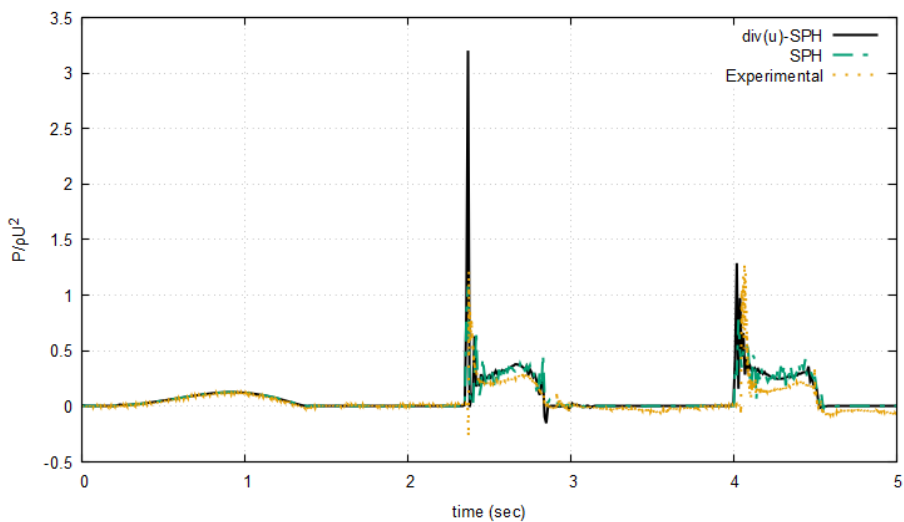

Figure 19. Temporal pressure comparison with GC #10 experimental data for *div(u)* and SPH.

## 5. Conclusions

In this paper, we introduced a novel divergence cleaning method for weakly compressible smoothed particle hydrodynamics (SPH) to mitigate the presence of non-physical acoustic waves in the numerical pressure field. Our approach employs a constrained hyperbolic (hyperbolic – parabolic) formulation to enhance the accuracy and stability of the SPH simulations. Through a series of benchmark tests, including Taylor-Green vortices, patch tests, jet impingement, and sloshing tank scenarios, we demonstrated that the proposed numerical scheme can effectively reduce acoustic oscillations while maintaining the physical accuracy of the pressure and velocity fields. The results show significant improvements in pressure smoothness and a substantial reduction in velocity divergence errors compared to the classical weakly compressible SPH formulation. The success of our method in various test cases highlights its potential for broader applications in simulations involving complex fluid-structure interactions and multiphase flows. Future work will focus on exploring its performance when adopted to more complex 3-D simulations.

## Acknowledgements

The authors would like to acknowledge the assistance given by Research IT and the use of the Computational Shared Facility at the University of Manchester. The first and third authors would like to acknowledge the UKRI Engineering and Physical Sciences Research Council (EPSRC), for the grant EP/W00755X/1 and EP/V039946/1. The second author would like to acknowledge the financial support from PNRR MUR project ECS_00000033_ECOSISTER and by University of Parma through the action Bando di Ateneo 2021 per la ricerca co-funded by MUR-Italian Ministry of Universities and Research - D.M. 737/2021 - PNR - PNRR – NextGenerationEU, CUP D91B21005370003